\documentclass[letterpaper,superscriptaddress,aps,pra,nolongbibliography,twocolumn,showpacs,floatfix]{revtex4-1} 

\usepackage{graphicx,helvet}
\usepackage{color}
\usepackage{bm}                      
\usepackage{tikz}
\usepackage{amsfonts}
 \usepackage{amsmath}
\usepackage{lipsum}
\definecolor{mygray}{gray}{0.4}
\definecolor{light-blue}{rgb}{0.8,0.85,1}
\usepackage{amsmath}%
\usepackage{MnSymbol}%
\usepackage{wasysym}%
\usetikzlibrary{decorations.shapes}
\renewcommand{\>}{\rangle}

\newcommand{\<}{\langle}

\newcommand{\mcM}{\mathcal{M}}
\newcommand{\mcN}{\mathcal{N}}
\newcommand{\mcQ}{\mathcal{Q}}

\newcommand{\mcC}{\mathcal{C}}
\newcommand{\mcK}{\mathcal{K}}

\newcommand{\cut}{\mathrm{cut}}
\newcommand{\rmd}{\mathrm{d}}

\newcommand{\rmf}{f}
\newcommand{\eref}[1]{eq.~(\ref{#1})} 

\newcommand{\fref}[1]{fig.~\ref{#1}}

\newcommand{\Eref}[1]{Eq.~(\ref{#1})} 
\newcommand{\Sref}[1]{Sec.~\ref{#1}}
\newcommand{\Fref}[1]{Fig.~\ref{#1}}  
\newcommand{\Tref}[1]{Table~\ref{#1}}

\newcommand{\tr}{\mathop{\mathrm{tr}}\nolimits}

\newcommand{\qv}{QV}
\newcommand{\old}{\infty}
\newcommand{\Mmed}{\mcM^{\<\cdot\>}}

\newcommand{\unam}{Universidad Nacional Aut\'onoma de M\'exico, M\'exico}
\newcommand{\ifunam}{Instituto de F\'{\i}sica, \unam}
\newcommand{\fisguadalajara}{Departamento de F\'isica, Universidad de Guadalajara, Guadalajara, Jal\'isco, M\'exico}
\newcommand{\ifimar}{Instituto de Investigaciones F\'isicas de Mar del Plata (IFIMAR), CONICET-UNMdP,  Mar del Plata,  Argentina}
\newcommand{\conicet}{Consejo Nacional de Investigaciones Cient\'ificas y Tecnol\'ogicas (CONICET), Argentina}

\newcommand{\uba}{Departamento de F\'isica ``J. J. Giambiagi'' and IFIBA, FCEyN, Universidad de Buenos Aires, 1428 Buenos Aires, Argentina}

\begin{document}
\title{Measuring and using non-markovianity}
\author{Carlos Pineda} \affiliation{\ifunam}
\author{Thomas Gorin} \affiliation{\fisguadalajara}
\author{David Davalos} \affiliation{\ifunam}
\author{Diego Wisniacki} \affiliation{\uba}
\author{Ignacio Garc\'ia-Mata} \affiliation{\ifimar} \affiliation{\conicet}
\email{carlospgmat03@gmail.com}

\begin{abstract}
We construct measures for the non-Markovianity of quantum evolution with a
physically meaningful interpretation.
We first
provide a general setting in the framework of channel capacities and propose
two families of meaningful quantitative measures, based on the largest revival
of a channel capacity, avoiding some drawbacks of other non-Markovianity
measures. We relate the proposed measures to the task of information screening. 
This shows that the non-Markovianity of a quantum process may be used as a 
resource. Under these considerations, we analyze two paradigmatic examples, a 
qubit in a  quantum environment with classically mixed dynamics
and the Jaynes-Cummings model.
\end{abstract}

%
%
\maketitle
\section{Introduction} 

The field of open quantum systems is of paramount importance in quantum theory
\cite{breuer2007theory}. It helps to understand fundamental problems as
decoherence, the quantum to classical transition, or the measurement problem
\cite{decoherencetheory}. Besides, it has been essential to reach an impressive
level of control in experiments of different quantum systems, which are the
cornerstone in recent development of quantum technologies
\cite{Haroche2001,Blatt2003,Bloch2008,photon-tech,Ladd:2010gb}.
 
The usual approach to quantum open systems relies on the assumption that the
evolution has negligible memory effects. This supposition is part of the
so-called Born-Markov approximation, which also assumes weak
system-environment coupling and a large environment.  The keystone of Markovian
quantum dynamics is the Lindblad master equation \cite{Lin76,GoKoSu76} which
describes the generator of quantum dynamical semigroups.  The
behavior of several interesting and realistic quantum systems has been studied
using the Born-Markov approximation.  However, these assumptions (weak coupling
or large size of the  environment) cannot  be applied in many situations,
including recent experiments of quantum control. This  shows the importance of
understanding quantum open systems beyond the Born-Markov approximation.

A great amount of work (see \cite{Breuer12,RHP14} and references therein) has
been done to understand and characterize non-Markovian quantum evolutions or
non-Markovianity (NM) -- as it is generically called.  This not only gives us a
better understanding of open quantum systems but also provides more efficient
ways to control quantum systems. For example, it was recently shown that
non-Markovianity is an 
essential resource in some instances of steady state
entanglement preparation~\cite{PhysRevLett.108.160402,Cormick2013} 
or can be
exploited to
carry out quantum control tasks that could not be realized in closed systems
\cite{nm-qc}. Besides, non-Markovian environments can speed 
up quantum evolutions reducing the quantum speed limit \cite{DeffnerLutz2013}.

Unlike other properties, like entanglement, there is not a unique definition
of non-Markovianity.  There exist different criteria, more or less physically
motivated, which in turn can be associated to a measure~\cite{Breuer12,RHP14}.
The two most popular criteria are based on 
distinguishability~\cite{PhysRevLett.103.210401} and 
divisibility~\cite{Rivas2010}, from which two corresponding measures can be 
derived. There exist other measures~\cite{Xiao2010,VasileManisc2011,Shunlong2012,Lorenzo2013,sabrinaScienRepot} 
which are basically variations of these two or are very similar. All of these 
measures present some of the following problems: Lack of a clear and intuitive 
physical interpretation, they can diverge in very generic 
cases~\cite{Znidaric2011}, and they are not directly comparable between them. 
Another problem is that, even if at least one of them has an intuitive physical 
interpretation, in terms of information flow~\cite{PhysRevLett.103.210401}, 
neither, to our knowledge,  has a direct relation to a resource associated to a
task -- like 
entanglement of formation has.

In this work, we pursue two goals. First, we want to construct NM measures
without the mentioned  drawbacks.  We undertake this task within the framework
of channel capacities. The proposed measures are based on the maximum revival
of the capacities, a characteristic that has a very simple physical
interpretation and has a natural time-independent bound.  Of course there might
-- and most likely will -- exist many possible measures of quantum non-
Markovianity.  Thus, we first provide a general setting and then put forward
two plausible, meaningful quantitative measures. Our second goal is to outline  the theoretical bases for considering NM as a resource.   

Consider what we call a quantum vault 
(\qv). 
Alice shall deposit information, classical or quantum, in a quantum 
physical system (say in a physical realization of a qubit); during some time,
through which the system evolves, the physical system can be subject to an 
attack by an eavesdropper, Eve; finally, after that time interval, the 
information is to be retrieved by Alice from the same physical system. Of 
course, the system interacts with an environment, to which neither Alice nor 
Eve can access. Notice that this task can be related with quantum data 
hiding~\cite{Divincenzo2002,Divincenzo2003,Mathews2009}.
We show that one of the NM measures proposed is closely related 
to the efficiency of the quantum vault. Therefore the value of the measure
can be considered a resource associated to a specific task.

To illustrate our ideas we analyze two examples of physical systems
coupled to non-Markovian environments, and analyze the newly defined measures as
well as their  quantum vault capabilities. We also explain possible
advantages with respect to other NM measures.  First we study a qubit, coupled
via pure dephasing to an environment whose dynamics are given by a mixed quantum
map.  Different kinds of dynamics can be explored, changing the initial state
of the environment. For the measures proposed in previous works, this sometimes
leads to an unexpected behavior.  The other example we consider is the well
known Jaynes-Cummings model (JCM) \cite{JaynesCumm1963}, a two level atom
coupled to a bosonic bath, where we contrast our proposals, with some of the
most used NM measures.   

The work is organized as follows. In Sec. \ref{measures} we describe the
general framework that relates NM with capacities of a quantum channel. Then, we
define two NM measures based on the largest revival of the capacities. In Sec.
\ref{sec:quantum_vault}
we introduce the concept of quantum vault and show its relation to 
the new NM measures.  \Sref{sec:examples}
is devoted to analyzing two examples using the ideas
presented in the previous sections.  We end the paper with some final remarks
in Sec. V.

\section{\label{measures} Non-Markovianity measured by the largest revival} 
The two most widely spread NM measures are, one due to Breuer et 
al.~\cite{PhysRevLett.103.210401} based on distinguishability
(in the following abbreviated by ``BLP''), and the other due to Rivas et
al.~\cite{Rivas2010} based on divisibility (``RHP''). At the heart of both 
measures, there is a well defined concept which has been borrowed from 
classical stochastic systems. In the case of BLP, it is the contraction of 
probability space under Markovian stochastic processes, while in the case of 
RHP, it is the divisibility of the process itself. Both concepts can be used as 
criteria for quantum Markovianity by {\it defining} that a quantum process is 
Markovian if the distinguishability between all pairs of evolving states is 
non-increasing (BLP-property), or if the process is divisible (RHP-property); 
otherwise the process is called non-Markovian. It has been shown in 
Ref.~\cite{RHP14}, that the semigroup property of a quantum process implies the 
RHP-property and that the RHP-property implies the BLP-property. In order to 
obtain a non-Markovianity (NM) measure, both groups of authors apply 
essentially the same procedure: integrate a differential measure for the
violation of the corresponding criterion. The same construction principle has
been used in Ref.~\cite{sabrinaScienRepot}, to quantify NM based on channel
capacities.

Consider the convex space of all quantum channels, and in this space a
continuous curve $\Lambda_t$ with $0 \le t \le \infty$ starting at the identity
$\Lambda_0 = \openone$. We will
call such a curve a {\it quantum process}. Any resource $\mathcal{K}$ of
interest will be a function on the space of quantum
channels. Thus any quantum process comes along with the function
\begin{equation}
K(t)= \mathcal{K}(\Lambda_t) ,
\label{eq:k:of:t}
\end{equation}
quantifying the resource the quantum channel provides at time $t$. Postulating
that $K(t)$ cannot increase during Markovian dynamics, one defines
the function
\begin{equation}
\mcM^\old_\mcK[\Lambda_t ]=
\int_{\dot K >0} \dot K(\tau)\, \rmd\tau \; ,
\label{} \end{equation}
as a measure for non-Markovianity.  
We use the subscript $\infty$ for this class of measures, as it is possible
that one has to add an infinite number of contributions (all intervals where
$\dot K$ > 0).
We use brackets to indicate a
functional on the space of quantum processes, and parenthesis when we refer to
a functional on the space of quantum channels. 
One can immediately derive a criterion for 
non-Markovianity, namely $\mcM^\old_\mcK[\Lambda_t ] > 0$. In the case of RHP,
\begin{equation}
\dot K(\tau)= \lim_{\varepsilon\to 0^+} 
   \frac{{\rm tr}|C_{\tau+\varepsilon,\tau}| - 1}{\varepsilon}\; ,
\end{equation}
where ${\rm tr}|\hat A|$ is the trace-norm and $C_{\tau+\varepsilon,\tau}$ is 
the Choi representation~\cite{BenZyc06,HeiZim12} of the map 
$\Lambda_{\tau+\varepsilon} \circ \Lambda_{\tau}^{-1}$, which evolves states 
from time $\tau$ to time $\tau+\varepsilon$. 
In the case of BLP
\begin{equation}
K(t)= D\big [\, (\Lambda_t(\rho_1), \Lambda_t(\rho_2)\, \big ] \; ,
\label{Def:BLPmeasure}\end{equation}
where $D(\varrho_1,\varrho_2)= {\rm tr}|\varrho_1 - \varrho_2|/2$ is the trace
distance between the two states $\varrho_1$ and $\varrho_2$. The initial states
$\rho_1,\rho_2$ are chosen to maximize the respective NM measure.
The measures defined in Ref.~\cite{sabrinaScienRepot},
can also be cast in this form. In that case, $K(t)$ is directly defined as
the corresponding channel capacity of $\Lambda_t$.

For definitiveness, two different channel capacities are considered in this
work, that for entanglement assisted communication, and that for quantum
communication~\cite{Smith2010}. Note that also much simpler measures
such as average fidelity, purity, or some measure of entanglement, may be cast
into that form.

This construction, which includes contributions from a possibly infinite number
of intervals where $\dot K > 0$, may result in rather inconvenient
properties. For instance, $\mcM^\old_\mcK[\Lambda_t]$ tends to overvalue small
fluctuations, which typically occur in the case of a finite environment,
finite statistics or experimental fluctuations. This
might even lead to the divergence of the measure. This can be remedied by
normalizing such as in~\cite{Rivas2010}, where the authors consider
$\mcM^\old_\mcK[\Lambda_t]\, (a + \mcM^\old_\mcK[\Lambda_t])^{-1}$ with $a=1$.
However, this normalization is completely arbitrary, as any other scale for $a$
would be equally well acceptable. Even if the measures yield finite values, it
is not clear how one should interpret a statement that one process has a larger
value for BLP-NM (RHP-NM) than another. It is even less possible to compare
values obtained for different measures. Therefore, these measures should be
considered as non-Markovianity criterion.

Here, we will show that a rather simple modification of the construction can
avoid these issues, and lead to a clear physical interpretation of the
resulting NM measures. The modification consists in considering only the
largest revival with respect to either (i) the minimum value of $K(\tau)$ prior
to the revival or (ii) to the average value prior to the revival.  Thus, we
take
\begin{equation}
\mcM^{\max}_\mcK[\Lambda_t ] = \max_{t_f,\tau \le t_f }
\left[ K(t_\rmf) - K(\tau) \right]
\label{eq:GorinNM}
\end{equation}
in the first case and
\begin{equation}
\mcM^{\<\cdot \>}_\mcK[\Lambda_t ] =
\max \left\{ 0,
\max_{t_f }
\left[
K(t_f) - \< K(\tau) \>_{\tau<t_f}
\right]
\right\}
\label{eq:CarlosNM}
\end{equation}
in the second. Here, $\< \cdot \>_{\tau<t_f}$ denotes time
average until $t_f$. 
In the first case, we are measuring the biggest revival during the time
interval whereas in the second, we are measuring a revival,
but with respect to the average behavior prior to this time.
Notice that
\begin{equation}
\mcM^{\<\cdot \>}_\mcK[\Lambda_t ] \le \mcM^{\max}_\mcK[\Lambda_t ]
\label{}
\label{eq:ineq}\end{equation}
as $\< K(\Lambda_{\tau}) \>_{\tau<\tau_{\rm max}} \ge \min_{\tau<\tau_{\rm max}}
K(\Lambda_{\tau})$.
Moreover, also notice that
\begin{equation}
\mcM^{\old}_\mcK[\Lambda_t ]>0
\iff
\mcM^{\max}_\mcK[\Lambda_t ] >0,
\label{eq:iff}\end{equation}
though no such relation is found for $\mcM^{\<\cdot \>}_\mcK$. In fact,
we shall see later that non-monotonic behavior {\it does not} guarantee
a positive value for $\mcM^{\<\cdot \>}_\mcK$.


\section{Non-Markovianity as a resource: quantum vault} 
\label{sec:quantum_vault}

We consider a quantum system, which is used to store and retrieve information by
state preparation and measurement. The quantum system is coupled to an
inaccessible environment and we describe its dynamics
by a quantum process. In order to use the system, Alice encodes her information
(which may be quantum or classical) in a quantum state. Then, at some later
time Alice attempts to retrieves the information from the evolved quantum
state.
Note that this state need not be equal to the initial state, it is sufficient
that Alice is able to recover her information from it.
The capacity of the device depends
on the amount of information which can be stored and faithfully retrieved.
During the time in which the information is stored, it might be subject to
an attack by an eavesdropper Eve. Some important remarks should be mentioned.
Eve has a finite probability of attacking, and her attack
destroys the quantum state.
We assume that Alice becomes aware if there is an attack and
discards the state.
A good  quantum vault is such that Alice can obtain her
information with high reliability and between state preparation
and read-out, the information is difficult to
retrieve.

The process, until the measurement by Alice or Eve, shall be described by the
quantum process $\Lambda_\tau$, while the information shall be quantified by a
capacity $\mcK$.  We shall thus have a time dependent value of the capacity,
analogous to \eref{eq:k:of:t}.  The times considered range from $0 \le \tau \le
t_\rmf $, with $t_\rmf$ being the time at which Alice attempts to retrieve the
information.
The average
information that can be obtained by Eve per attack is then $\< K \>$,
where the average is taken during the vault operation, namely from
$0$ until $t_\rmf$. Here we assume that when Eve attacks, she does 
only once, as an attack destroys the state anyway.
If Eve attacks with a probability $q$, on average she will obtain the
information $q \< K \>$.
Thus, the average information successfully retrieved
by Alice will be only $(1-q) K(t_\rmf)$. We shall consider as a figure of
merit the difference between the average information obtained by Alice, and
the one obtained by Eve:
\begin{equation}
\Delta K_q = (1-q) K(t_\rmf) - q \<  K \> .
\end{equation}
Note that this quantity can be negative, when Eve obtains on average more
information than Alice can retrieve. Finally, we may define
the quantum vault efficiency as $\eta_q = \Delta K_q /K_\text{max} $,
with $K_\text{max} = \mcK(\openone)$.
A good quantum vault should have an efficiency close to one.

Assume that $\mcK$ is normalized in such way that
$p = K/K_\text{max}$ is the probability that the message encoded in
the state will be retrieved.
A successful run can be defined as a run in which, if Eve attacks, she
gains no information, whereas if Eve does not attack, Alice retrieves
successfully the information. From the considerations above, one can see
that the probability of having a successful run is given by
\begin{equation}
P_q = q (1-\<p(t)\>) + (1-q)p(t_f) = \eta_q + q.
\label{eq:Pq}
\end{equation}
Associated with this probability, we shall associate a quality factor for the
channel, as a quantum vault, that is simply the above probability, weighted by
the capacity of the channel, that is $ \mcN_q = K_\text{max} P_q = \Delta K_q
+q K_{\max{}}$.

Now we shall discuss $\Delta K_q$ and $\mcN_q$ for some particular examples and
establish its relation to $\Mmed_\mcK$.  We first examine the worst case scenario:
$q\approx 1$. By definition, if Eve attacks she destroys the state. This fact
is reflected in $\Delta K_q$ which can go from minimal value $-K_{\text{max}}$
(worst efficiency $\eta_q$) to  $\Delta K_q=0$ (poor \qv{}), when $\<K(t)\>\approx
0$.  $\mcN_q$ on the other hand ranges from $\mcN_q=0$ (i.e. bad \qv{}) to
$\mcN_q=K_{\max{ }}$.
In the last case large $\mcN_q$ value due to small $\<K(t)\>$ evidences the fact
that Eve is unable to obtain anything.  In the best case scenario of $q\ll 1$,
evidently, the efficiency of the vault is only tied to $K(t_f)$, the larger the
better.

Now let us assess the general case. We will only take into account the case
where $\Mmed>0$, i.e. from \Eref{eq:CarlosNM} there is at least one $t_f$ for
which $\<K(t)\> < K(t_f)$.
The first relation between the \qv{} and $\Mmed_\mcK$ that we find  is
\begin{equation}
\text{min}(1-q,q)\times\Mmed_\mcK\le \Delta K_q \le \text{max}(1-q,q)\times\Mmed_\mcK,
\label{eq:DeltaKineq}
\end{equation}
which is easy to derive from the definition, provided $t_f$ is the same for
both.  This relation sets  lower and upper bounds for the \qv{}, depending on $q$
and $\Mmed$ [a corresponding relation with $\mcN_q$ follows directly from
\Eref{eq:Pq}].  A reasonable assumption is that information about the
probability of attack by Eve, $q$ is not known.  One can invoke a maximum
entropy principle to use the average value of $q$, namely $q=1/2$.  For this
unbiased case, (and $\Mmed>0$)  we have
\begin{equation}
\Delta K_{1/2}=\frac{1}{2}\Mmed_\mcK, \quad
\mcN_{1/2}=\frac{1}{2}(K_\text{max}+\Mmed_\mcK).
\label{eq:relation:resource:monotonicity}
\end{equation}
These equations relate the NM measure proposed, with the
possibility to perform the task at hand.  In particular, it gives an
operational meaning to the measures proposed here, and show that, for the task
proposed here, the figure of merit is $\Mmed_\mcK$.

\begin{figure} 
  \includegraphics[width=\linewidth]{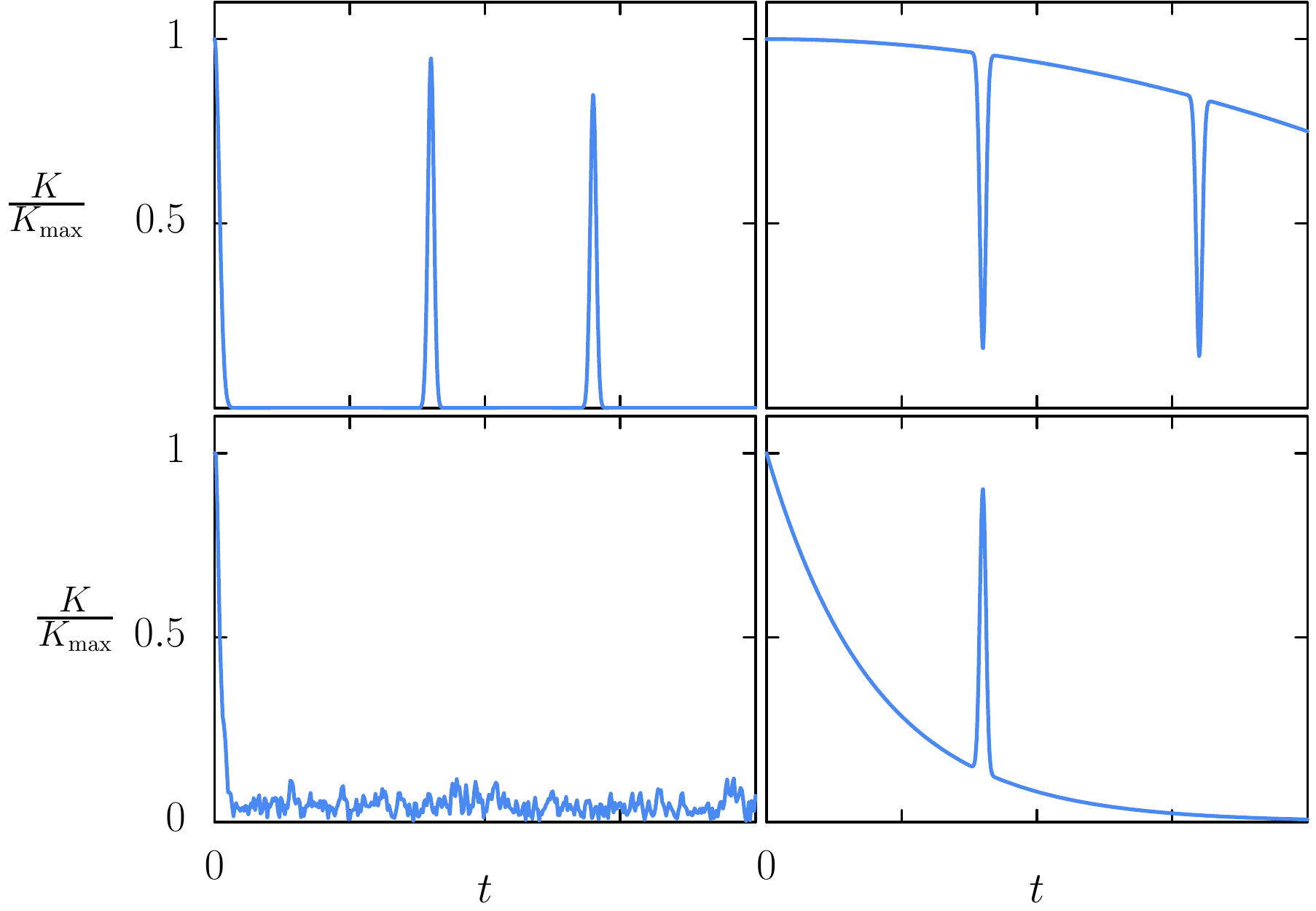} 	%
  \caption{Schematic examples of $K(t)/K_{\text{max}}$ to be considered as
  \qv{}. (top-left) Example with large
  $\delta K_q$ and $\Mmed$, good \qv{} candidate. (top-right) Worst case scenario. Small $\delta K_q$, small (or null) $\Mmed$
  and infromation always available for EVE to grab. (bottom-left) Small $\Mmed$ an $\Delta K_q$ (poor in for information retrieval)
  but decent $\mcN_q$ (good for information protection). (bottom-left) More general case,
  strongly depending on $\Mmed$ and the height of the peak at $t_f$.
  \label{fig:ejemplos}}
\end{figure}

In \Fref{fig:ejemplos} we show some examples one could encounter for
$K(t)/K_\text{max}$.  In the first one
(Fig. \ref{fig:ejemplos}, top-left) we have $\<K(t)\>\approx 0$, $\Mmed\approx
K(t_f)$, and
\begin{equation}
\Delta K_q\approx (1-q)\Mmed
\end{equation}
so for a fixed $q$, a large $\Mmed$ implies high efficiency.  This is in fact
the ideal scenario for a \qv{}, because most of the time the information is
hidden and inaccessible to Eve and at time $t_f$ the information can be
retrieved with high accuracy.  
If $\<K(t)\>$ is very large, close to $K_\text{max}$ (e.g. Fig.
\ref{fig:ejemplos} top-right), then, by definition, the channel is not a good
\qv{}: a large proportion of the information is readily available at all times
before $t_f$.  Here 
$K(t_f) < \<K\>$
(so \Eref{eq:DeltaKineq} does not hold),
but the only possibility to have good efficiency is the trivial $q\to 0$ case.
If, on the other hand,  $\<K(t)\>$ and $K(t_f)$ are  both very small
(\Fref{fig:ejemplos} bottom-left) -- again $\Mmed\approx 0$-- there is little
chance to retrieve the information, even for small $q$, yielding poor
efficiency and  $\mcN_q$.  
For large $q$, $\mcN_q\approx q K_\text{max}$ can be large.
The interpretation of this large value of $\mcN_q$ is that Eve
will likely attack, but unsucceesfully.
Here the interpretation of this large value of $\mcN_q$ is that Eve 
will likely attack, but unsucceesfully.
Finally, we consider the case where $K(t)$ decays monotonously except for
one bump (e.g. Fig. \ref{fig:ejemplos} bottom-right).  The analysis now
requires a little more care.   If $K(t_f)<\<K\>$, which happens for a small enough 
bump, then $\mcM^\cdot_\mcK=0$, and
there is no
connection between $\Delta K_q$ (or $\mcN_q$) and $\mcM^\cdot_\mcK$.  The
analysis is similar to that of Fig. \ref{fig:ejemplos} (top-right).  On the
other hand if
$\Mmed>0$ the efficiency is bounded by \Eref{eq:DeltaKineq}  and for maximum
$\Mmed$, the case is equivalent to the first one (top-left).

\section{Examples} 
\label{sec:examples}
In this section we present concrete physical examples of quantum channels where
we can test the newly proposed measures and
their relation with the \qv{} scheme.
\subsection{Environment with mixed dynamics} 
\newcommand{\opoo}{\op{0}{0}}
\newcommand{\opoi}{\op{0}{1}}
\newcommand{\opio}{\op{1}{0}}
\newcommand{\opii}{\op{1}{1}}
\newcommand{\ktimes}{\rangle\! \langle}
\newcommand{\op}[2]{|#1\ktimes #2|}
\newcommand{\e}{{\rm env}}
\newcommand{\s}{{\rm sys}}

Let us discuss encoding quantum information in a qubit coupled to an
environment in a dephasing manner.  We consider that the environment evolves
according to a dynamics that in the semiclassical limit is mixed, i.e. has
integrable and chaotic regions in phase space.

A simple way to realize such environment is using a controlled kicked 
quantum map~\cite{NachoCarlosDiego2012,
NachoCarlosDiego2014}.  In this case, the environment
evolution is slightly modified depending on the state of the qubit. This is
equivalent to having a coupling with the environment that commutes with the
Hamiltonians corresponding to the free evolution of  each part,
qubit and environment.

\begin{figure} 
  \includegraphics[width=\columnwidth]{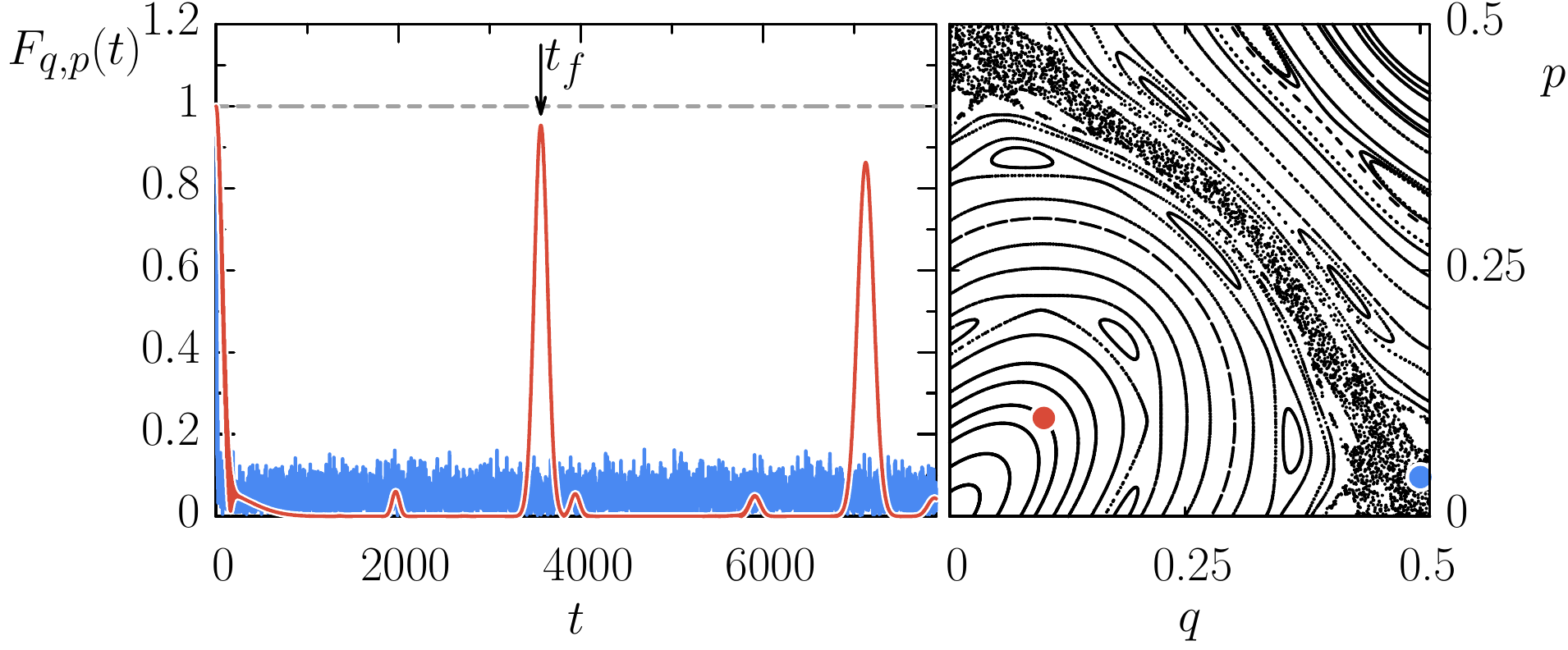}
  \caption{(left) $F(t)$ as a function of time for two different initial states
  of an environment modeled by the quantum Harper map. (right) the classical
  phase space of the environment for the parameters studied (see main text).
  The two states of the environment in the left figure are coherent pure states
  centered where the color dots are shown in the phase space portrait, i.e. one
  in the chaotic region (blue) and the other in the integrable region (red),
  corresponding to similar colors in the left panel.
  }
  \label{fig:peaked:fidelity}
\end{figure} 
Here we choose to use the quantum Harper map~\cite{leboeuf1990}.
The evolution operator, in terms of the discrete conjugate
space-momentum variables $\hat q$ and $\hat p$ is
\begin{equation}
U_{k}=
\exp\left[-i \frac{k}{\hbar} \cos(2 \pi \hat p) \right]
\exp\left[-i \frac{k}{\hbar} \cos(2 \pi \hat q) \right]
\label{harperU},
\end{equation}
$\hbar\equiv 1/(2\pi N)$ being the effective Planck constant and
$N$ the dimension of the Hilbert space of the environment.
The corresponding classical dynamics ($N\to \infty$) is given by
\begin{align}
p_{n+1}&=p_n-k\sin(2 \pi q_n),\nonumber \\
q_{n+1}&=q_n+k\sin(2 \pi p_{n+1}).
\end{align}
The phase space geometry is a 2-torus, so $p_n$ and  $q_n$ are taken modulo 1.
For $k=0.2$, the dynamics is mixed, see Fig.~\ref{fig:peaked:fidelity}.
To use this closed system as an environment, we consider that the state of the
qubit induces a small change in the parameter $k$ of the map, so the evolution
of the whole system for one time step is given by the Floquet operator
\begin{equation}
U=\opoo U_{k}+\opii U_{k+\delta k}
\label{eq:Us}
\end{equation}
and $U(t) = U^t$, for integer  $t$. Throughout this example, we shall
set $N=4000$, $k=0.2$ and $\delta k=2 \hbar$, unless otherwise stated.
The initial state of the whole system is the uncorrelated state
$\rho_\s \otimes \rho_{\e,q,p}$, where the
environment will be taken to be a pure coherent state centered in $(q,p)$.
The state of the qubit, obtained with unitary evolution in the whole system
and partial tracing the environment, is given by
\begin{equation}
\label{eq:rhot}
\rho_\s(t) =\tr_\e \left[U(t)\rho_\s\otimes \rho_{\e,q,p} U^\dagger(t)\right]
=\Lambda^{q,p}_t(\rho_\s).
\end{equation}
In the basis of Pauli matrices $\{\sigma_i\} =\{
\mathbb{I},\sigma_{x},\sigma_ y,\sigma_z\}/\sqrt{2}$,
the induced channel takes the form
\begin{equation}
\Lambda_{q,p}(t) =\begin{pmatrix}
1 & 0 & 0 & 0 \\
0& {\rm Re}[f_{q,p}(t)]& {\rm Im}[f_{q,p}(t)]&0\\
0& {\rm Im}[f_{q,p}(t)]& {\rm Re}[f_{q,p}(t)] &0 \\
0 & 0 & 0 & 1
\end{pmatrix},
\label{eq:paudepha}
\end{equation}
where
\begin{equation}
f_{q,p}(t)= \tr\left[ \rho_{\e,q,p} U_{k+\delta k}(t)^\dagger U_k(t)\right]
\end{equation}
is the expectation value of the echo operator $U_{k+\delta k}(t)^\dagger
U_k(t)$ with respect to the corresponding coherent state (also known as
fidelity amplitude).  We shall also define the fidelity,
$F_{q,p}(t)=|f_{q,p}(t)|$, for later convenience. Notice that the channel
depends, up to unitary operations, only on $F_{q,p}(t)$, and thus all
capacities will be functions solely of this quantity.

In  \Fref{fig:peaked:fidelity}~(left) we show two examples of $F_{q,p}(t)$, for
different initial conditions of the environment (marked with circles in
\Fref{fig:peaked:fidelity} (right). One can see that in case the
environment starts inside the chaotic sea, the system has very small
$\mcM^{\<\cdot\>}_F$ and $\mcM^{\text{max}}_F$ and therefore from
\Eref{eq:relation:resource:monotonicity}, it will be a very bad \qv{}.

\begin{figure} 
  \includegraphics[width=\columnwidth ]{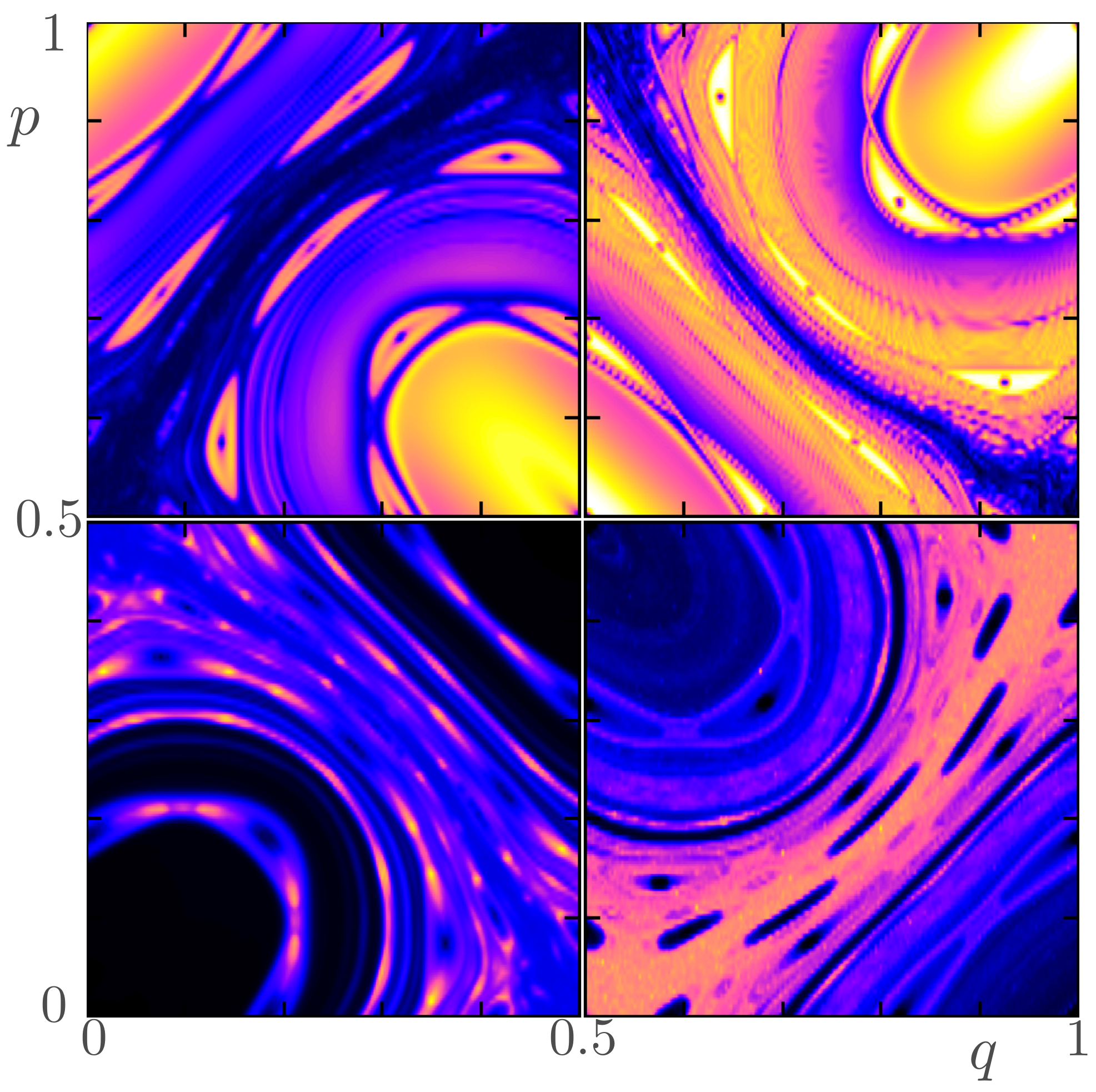} %
  \caption{(Color online)
	Mapping of classical phase space obtained from the different
	non-Markovianity measures discussed, for the quantum Harper map with
	$N=8000$, $K=0.2$, $\delta K/\hbar=2$ and maximum time $t_{\cut}=16000$.
        The color code is dark/black$=0$,  light/white$=$ max. 
	The different subfigures correspond to 
	$\mcM^{\<\cdot\>}_F$, top left with a maximum value of the measure
	of $1$; $\mcM^{\text{max}}_F$ on the top right with maximum value of $1$;
	$\mcM^{\rm BLP}$, bottom left, with a maximum value of $400$;
  	and at the bottom right, $\mcM^{\rm RHP}$ with a maximum value 
	of $19000$. 
  }
  \label{fig:phase:space}
\end{figure}

An interesting point of \Fref{fig:peaked:fidelity} is how
$\mcM^{\<\cdot\>}_F$ and $\mcM^{\text{max}}_F$  compare to
$\mcM^{\text{BLP}}$ and $\mcM^{\text{RHP}}$.
\begin{table} 
     \begin{ruledtabular}
     \begin{tabular}{ccccc}
     & $\mcM^{\<\cdot\>}_F$&$\mcM^{\rm max}_F$ &$\mcM^{\rm BLP}$ & $\mcM^{\rm RHP}$\\
    \hline
     {\color{red} $\bullet$} & 0.899 & 0.953 & 6.6 & 1333.97     \\
     {\color{blue} $\bullet$}& 0.108 & 0.194 & 125.84 & 6274.89
     \end{tabular}
     \end{ruledtabular}
     \caption{Comparison between different values of measures of
     non-markovianity for the two situations depicted in
     \fref{fig:peaked:fidelity}, with corresponding colors. We cut the integral
     in \eref{eq:BLPRHPF} in $t=8000$.  The inherent fluctuations present for
     this finite dimensional environment cause the integrable situation (with
     larger fluctuations) to reach larger values for the NM measures $\mcM^{\rm
     BLP}$ and $\mcM^{\rm RHP}$, than the chaotic counterpart. On the other
     hand, both $\mcM^{\<\cdot\>}_F$ and $\mcM^{\rm max}_F$ capture well the
     idea of \qv{}, reporting large values for the integrable case, and small
     values for the chaotic one.
     \label{table}}
\end{table} 
In terms of $F_{q,p}(t)$ the latter are given by
\begin{align}
  \mcM^{\text{BLP}}&=2 \int_{0, \dot{F}>0}^{t_\cut} d\tau \dot{F}_{q,p}(\tau), \nonumber \\
  \mcM^{\text{RHP}}&=\int_{0, \dot{F}>0}^{t_\cut} d\tau\frac{\dot{F}_{q,p}(\tau)}{F_{q,p}(\tau)},
\label{eq:BLPRHPF}
\end{align}
where $t_\cut$ indicates a cutoff time.
The $[F_{q,p}(\tau)]^{-1}$ term in $\mcM^{\text{RHP}}$ can be problematic when
$F_{q,p}(\tau)$ is very small, which is exactly the case for an initial state
located in the chaotic region. In \Tref{table}
the values of all four
measures, based on $F$, corresponding to \Fref{fig:peaked:fidelity} are shown.
The values reported in the table highlight important characteristics of all four measures.
On the one hand, for non-monotonicity based measures, intuitively we expect that
a fast decaying $K(t)$ followed by sharp, and high revivals would yield a larger
value of non-Markovianity. This is not the case for $\mcM^{\text{BLP}}$ and $\mcM^{\text{RHP}}$
(at least in this particular example), for different reasons. In the case
of $\mcM^{\text{RHP}}$ it is due to the small denominator and in the case of
$\mcM^{\text{BLP}}$ it is due to fluctuations (and finite $N$).
These facts are further illustrated in the color density plots of 
\Fref{fig:phase:space}. We see that in all cases the underlying classical 
structure is clearly outlined. For both $\mcM^{\<\cdot\>}_F$ and 
$\mcM^{\text{max}}_F$ an additional structure appears that resembles the 
unstable manifolds. The measures $\mcM^{\<\cdot\>}_F$, 
$\mcM^{\text{max}}_F$, and $\mcM^{\text{BLP}}$ all seem to peak in the
vicinity of the border between chaotic and regular behavior. As stated before,
the $\mcM^{\text{RHP}}$ measure behaves differently, as it is larger in the 
chaotic region.

Notice that the measure $M^{\<\cdot \>}_F$ can be associated with 
the task of transmitting classical information (without the use of
entanglement) encoded initially in the states $|\pm \>$.

\subsection{Non markovian Jaynes-Cummings model} 
In order to explore and compare the different measures of non-Markovianity
discussed throughout this work, we now consider the paradigmatic
Jaynes-Cummings model~\cite{JaynesCumm1963}, which has served 
as testbed in quantum optics; see e.g.~\cite{breuer2007theory}. In this model, 
a two level atom is coupled to a bosonic bath, which induces a degradable 
channel in the qubit. We shall take advantage of the fact that a lot is known 
about this model analytically and we will build upon known results. 

The Hamiltonian of the system is $H=H_0+H_I$, where $H_0$ is the free 
Hamiltonian of the atom plus the reservoir and $H_I$ the interaction between 
them. In particular, 
$H_0 = \omega_0 \sigma_+ \sigma_- + \sum_k \omega_k b_k^{\dagger} b_k$, where 
$\sigma_\pm$ are the rising and lowering operators in the atom, $\omega_0$ is 
the energy difference between the  two levels in the atom, $b_k$ and 
$b_k^\dagger$ are creation and annihilation operators of mode $k$ of the bath, 
and $\omega_k$ its frequency. The interaction Hamiltonian is given by
$H_I = \sigma_+ \otimes B + \sigma_- \otimes B^{\dagger}$,  with 
$B=\sum_k g_k b_k$ and $g_k$ the coupling of the qubit to mode $k$. In the 
limit of an infinite number of reservoir oscillators and a smooth spectral 
density, this model leads to the following channel~\cite{breuer2007theory}:
\begin{equation}
\Lambda_t \left[\rho\right] = \left( \begin{array}{cc}
1-\vert G(t) \vert^2\rho_{ee}   & G(t) \rho_{ge} \\ 
G^{*}(t)\rho_{ge}^* &  \vert G(t) \vert^2 \rho_{ee}
\end{array}  \right),
\end{equation}
where the initial state
$
\rho = \begin{pmatrix} 1-\rho_{ee}& \rho_{ge} \\ 
\rho_{ge}^* &\rho_{ee} \\ 
\end{pmatrix}$.
\begin{figure}
\centering
\includegraphics[]{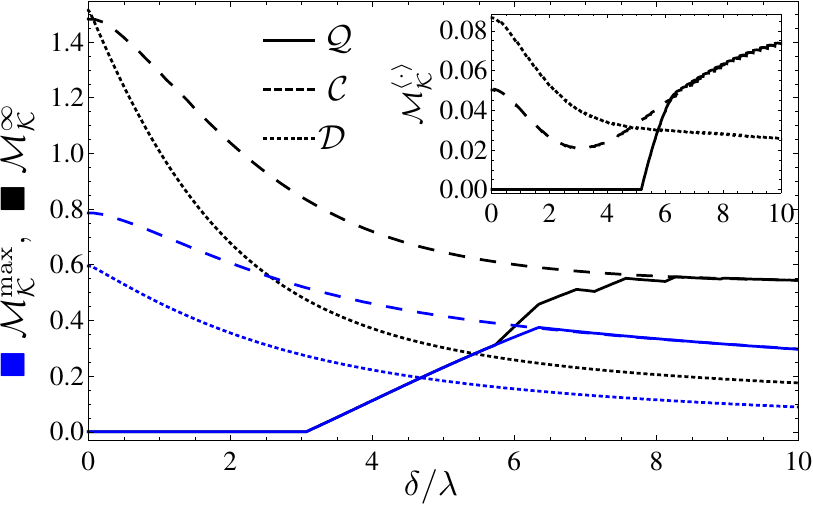} 
\caption{Comparative results of the measures of quantum non Markovianity treated in this work for the Jaynes-Cummings model, as a function of the scaled detuning $\delta/\lambda$, with a coupling of 
$\gamma= 20 \lambda$. We show three different types of measures, 
$\mcM^\old_\mcK$ (black lines), $\mcM^{\max{ }}_\mcK$ (blue lines), and 
$\mcM^{\< \cdot \>}_\mcK$ (inset) for three different capacities: Quantum capacity
$\mcQ$ (solid lines), classical entanglement assisted capacity $\mcC$ (dashed 
lines), and the capacity based on distinguishability $\mathcal{D}$ (dotted 
lines).
\label{jmmany}}
\end{figure}
The function $G(t)$ is the solution to the equation 
$\dot{G}(t) = - \int_{0}^{t} d\tau f(t-\tau) G(\tau)$, with $G(0)=1$, and 
$f(t-\tau)$ is the two-point correlation function of the reservoir. For a 
Lorenzian spectral density
\begin{equation}
f(t) = \frac{1}{2} \gamma_0 \lambda e^{-\vert t \vert \left( \lambda -i \delta\right)}, 
\end{equation}
we find 
find
\begin{equation}
G(t)= 
e^{-\frac{1}{2} t (\lambda -i \delta)} 
\left[
\frac{(\lambda -i \delta)}{\Omega} \sinh \left(\frac{\Omega t}{2} \right)
+\cosh \left(\frac{ \Omega t}{2} \right)
\right],
\label{G}
\end{equation}
where $\Omega=\sqrt{-2 \gamma  \lambda +(\lambda -i \delta)^2}$. 
Here, $\gamma_0$ is the strength of the system-reservoir coupling, $\lambda$ is
the spectral width, and $\delta$ is the detuning between the peak frequency of 
the spectral density and the transition frequency of the atom~\footnote{The
case without detuning is treated in detail in~\cite{breuer2007theory}; the
present case with detuning has first been 
solved in Ref.~\cite{sabrinaScienRepot} but with a minor mistake. Here, we 
present the corrected expression.}.

In what follows, we study the NM measures $\mcM_\mcK^\old$, 
$\mcM_\mcK^{\max{}}$, and $\mcM_\mcK^{\<\cdot\>}$ for the capacities 
$\mcQ$ (quantum capacity), $\mcC$ (entanglement-assisted classical 
capacity), and $\mathcal{D}$ (distinguishability of the states $|\pm\>$). 
The quantum capacity is defined as the maximal amount of quantum information 
(per channel use, measured as the number of qubits) that can be reliably 
transmitted through the channel.  It is given explicitly in terms of the
following maximization~\cite{Giovannetti2005}: $\max_{p \in [0,1]} \lbrace
H_2\left(\vert G(t) \vert^2 p\right) -H_2\left((1-\vert G(t) \vert^2) p \right)
\rbrace$ for $\vert G(t) \vert^2>1/2$ and 0 for $\vert G(t) \vert^2\leq1/2$.
The entanglement-assisted classical capacity $\mcC$ is defined as the maximal
amount of classical information (per channel use, measured as the number of
classical bits) that can be reliably transmitted through the channel, when 
Alice and Bob are allowed to use an arbitrary number of shared entangled 
states~\cite{Smith2010}. For the present channel it is given by~\cite{Giovannetti2005}:
$\mcC = \max_{p \in [0,1]} \lbrace H_2(p)+H_2\left(\vert G(t) \vert^2 p\right)
-H_2\left((1-\vert G(t) \vert^2) p \right) \rbrace$. Finally, we also consider 
the BLP measure defined in Eq.~(\ref{Def:BLPmeasure}). In this case the initial 
states which maximize the various types of NM measures may be chosen invariably
as the two eigenstates of the Pauli matrix $\sigma_x$~\cite{Breuer12}. Thereby,
we obtain $K(t) = |G(t)|$.

\begin{figure} 
\centering
\begin{tikzpicture}
\node at (0,0) {\includegraphics[width=7 cm]{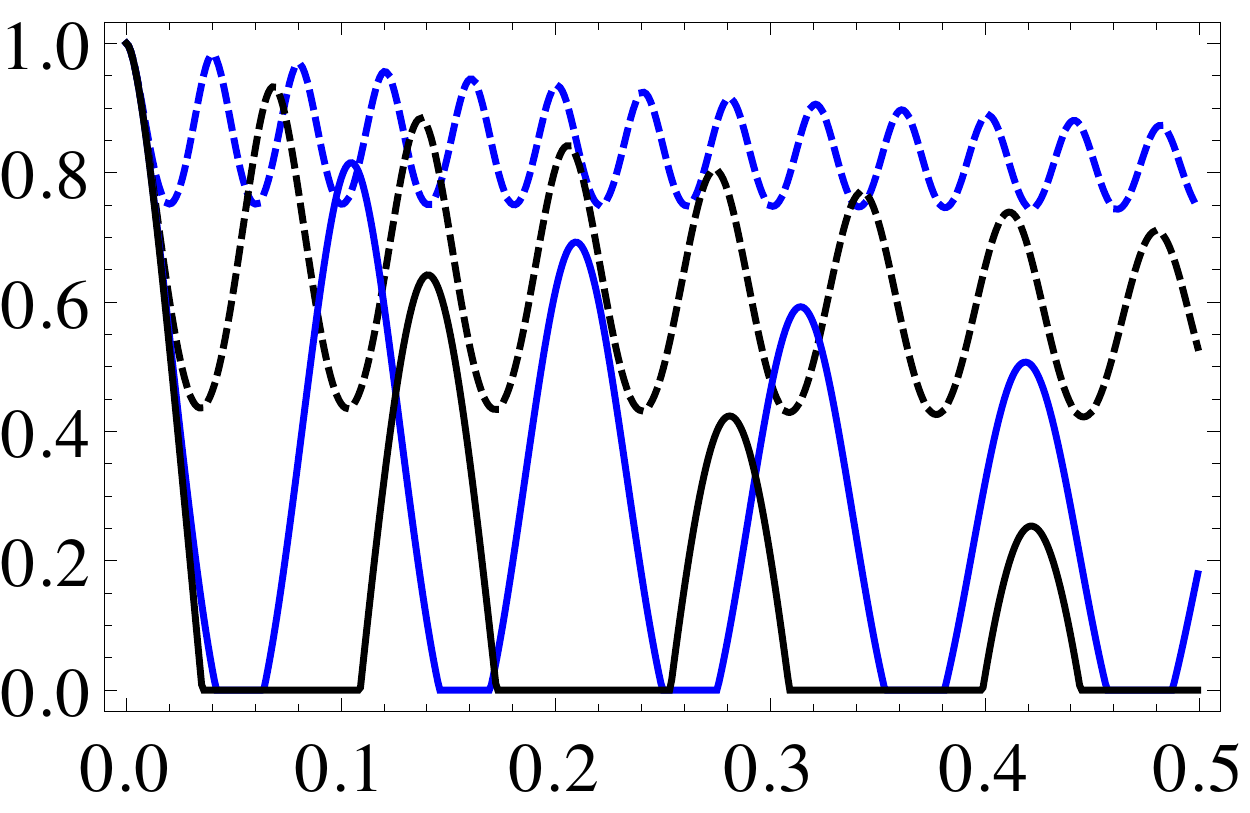}}; 
\node at (-4, 0) {\rotatebox{90}{$\mcQ$}};
\node at (0,-2.8) {$\lambda t$};
\end{tikzpicture}
\caption{Quantum capacities of the non markovian Jaynes-Cummings model with a strong coupling $\gamma/\lambda=1000$. The solid curves correspond to $\delta/\lambda=0$ (black), $\delta/\lambda=40$ (blue). The dashed curves are $\delta/\lambda=80$ (black), $\delta/\lambda=150$ (blue). \label{acap}}
\end{figure}
\begin{figure}  
\centering
\begin{tikzpicture}
\node at (0,0) {\includegraphics[width=6 cm]{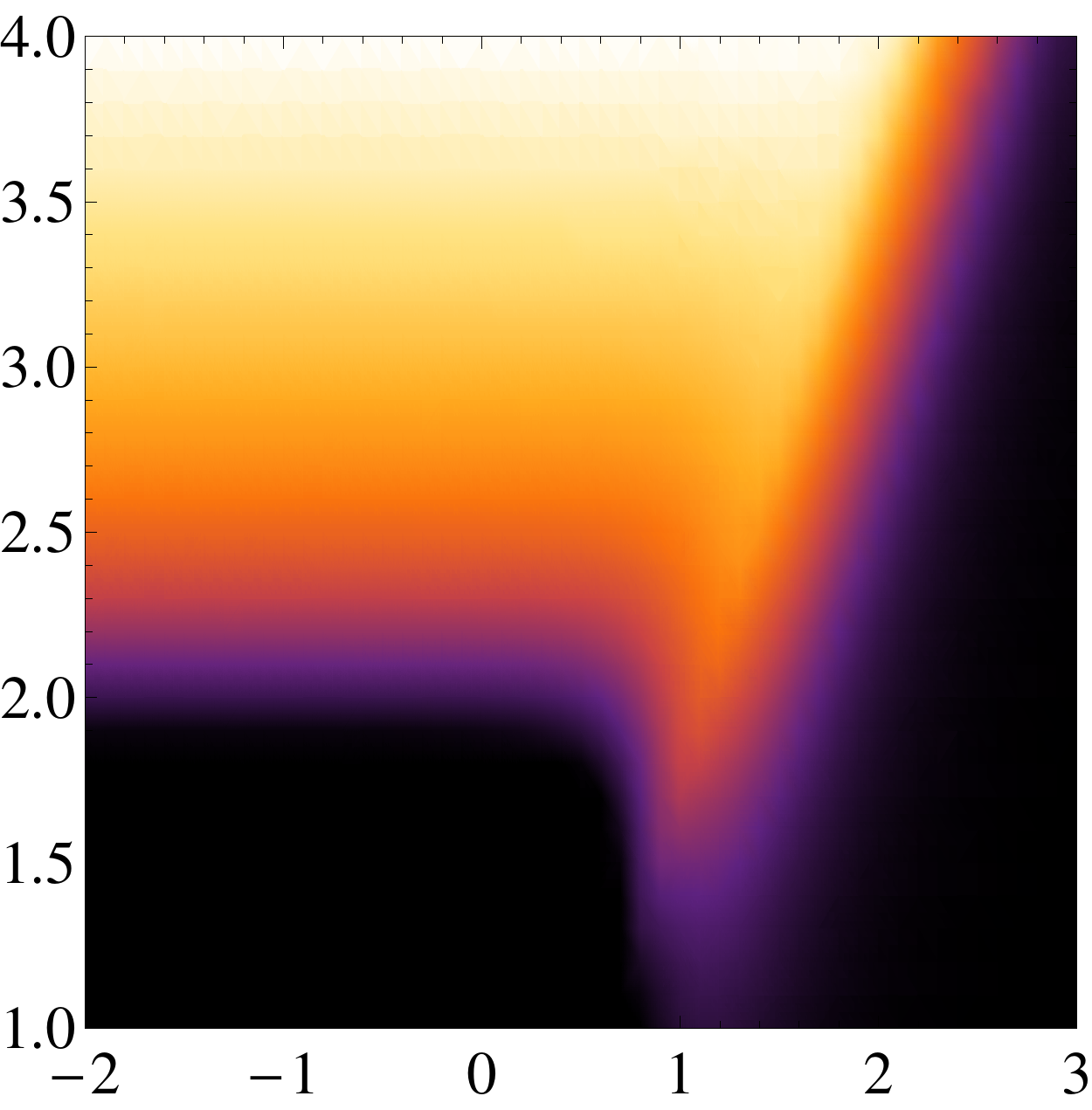}};
\node at (3.5,0) {\includegraphics[width=0.88 cm]{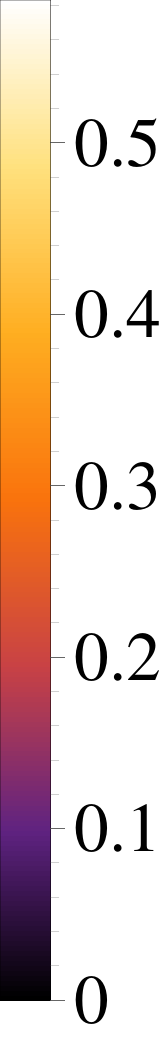}};
\node at (0,-3.4) {$\log_{10} \delta/\lambda$};
\node at (-3.4,0) {\rotatebox{90}{$\log_{10}\gamma/\lambda$}};
\end{tikzpicture}
\caption{Density plot of the $\mcM^{\langle \cdot \rangle}_\mcQ$ for the non
Markovian Jaynes-Cummings model as a function of its parameters.
\label{aden}}
\end{figure}

 
Figure~\ref{jmmany} shows a comparison between the measures 
$\mcM_\mcK^{\< \cdot \>}$ and $\mcM_\mcK^{\max{ }}$, introduced in this work,
and their counterpart $\mcM^\old_\mcK$. 
The measures regarding the average $\mcM_\mcK^{\< \cdot \>}$ are notoriously
smaller than the measures regarding both the maximum revival and the integrated
revivals. In fact, we find that 
$\mcM^{\< \cdot\>}_\mcK \le \mcM^{\max{ }}_\mcK$, in agreement with (\ref{eq:ineq}), and 
$\mcM^{\max{ }}_\mcK \le \mcM^\old_\mcK$.   
The measures related to the BLP criterion (dotted lines) behave similarly in 
all three cases, decaying monotonously with $\delta/\lambda$. In the case of 
the entanglement assisted classical capacity, $\mcM_\mcC^\old$ and 
$\mcM_\mcC^{\max{ }}$ also decay monotonously. However, 
$\mcM_\mcC^{\< \cdot\>}$ has a minimum at $\delta/\lambda \approx 3$. Beyond
that point $\mcM_\mcC^{\< \cdot\>}$ increases a bit further, but finally decays
to zero. In this region, $\mcM_\mcC^{\< \cdot\>} = \mcM_\mcQ^{\< \cdot\>}$
which is shown in Fig.~\ref{aden}.
The measures related to the quantum capacity (solid lines) show the most 
complicated behavior. $\mcM_\mcQ^\old$ and $\mcM_\mcQ^{\max{ }}$ are equal to
zero until $\delta/\lambda \approx 3$, $\mcM_\mcQ^{\< \cdot \>}$ is equal to
zero until $\delta/\lambda \approx 5$. Beyond these points, the measures
increase linearly. From $\delta/\lambda \approx 6.5$ on, $\mcM_\mcQ^{\max{ }}$ 
and $\mcM_\mcQ^{\< \cdot \>}$ reach the corresponding curves for the classical 
capacity. For somewhat larger values of $\delta/\lambda$ 
this also happens for the $\mcM_\mcQ^\old$. 

The fact that for quantum capacities, $\mcM_\mcQ^{\old}$ and 
$\mcM_\mcQ^{\max}$ start to deviate from zero at the same point 
$\delta/\lambda \approx 3$, illustrates \eref{eq:iff}. 
Other interesting feature that can be appreciated in $\mcM_\mcQ^{\old}$
are the multiple discontinuities in the derivative with respect to the
detuning. This is due to the sudden appearance of new bumps
in the quantum capacity of the channel, to which this measure is
sensible.  Moreover, for 
most instances  of $\mcK$, $\mcM_\mcK^{\old}$ can be discontinuous in the space 
of finite quantum processes, if we consider the maximum-norm, so the measures 
are not stable with respect to small deviations in the quantum dynamics. 
In particular, small amplitude high frequency noise can make $\mcM_\mcK^{\old}$
increase arbitrarily, whereas, it has a small effect (proportional to the
amplitude) in the case of $\mcM_\mcK^{\< \cdot \>}$ and $\mcM_\mcK^{\max{}}$. 

Regarding the classical capacity, it is worth noticing that the different
cases do not share the same tendency, $\mcM_\mcC^{\max{}}$ and
$\mcM_\mcC^{\old}$ diminishes with the detuning (as opposed to the quantum
capacity cases), but $\mcM_\mcC^{\< \cdot \>}$ has a non monotonic
behavior, mimicking fidelity until $\delta/\lambda \approx 3$, 
and then resembling the quantum capacity. A direct consequence 
inherited from the fact that $\mcQ \le \mcC$ is that
$\mcM_\mcQ^{\ \cdot } \le \mcM_\mcC^{\ \cdot }$ as can be seen
from the fact that, for all colors, the dashed line bounds 
the continuous line (here, the dot  denotes any 
of $\max{}$, $\< \cdot \>$ or $\old$). 
As a general remark, we also observe that $\mcM_\mcK^{\< \cdot \>}$
is much smaller than $\mcM_\mcK^{\max{ }}$ and $\mcM_\mcK^{\old}$. 
This is due to the fact that the peaks in the different capacities 
are thick, and the system, in fact, would not serve as a good 
quantum vault. 

Figure \ref{acap} shows the evolution of several quantum capacities for the 
Jaynes-Cummings model
varying the detuning, while keeping the reservoir coupling 
fixed. The table \ref{tableap} shows the values of the corresponding 
measures of non-Markovianity treated in this work. It shows that large 
detunings lead to poor scenarios for a quantum vault operation, while for zero 
and small detuning there are better situations for a usage of the \qv. The time 
when the first peak in the capacity appears can be tuned by choosing the 
correlation time of the bath. The figure \ref{aden} shows a density plot of the 
non-Markovianity measures, $\mcM^{\langle \cdot \rangle}_\mcQ$, as a function of 
the channel parameters $\delta$ and $\gamma$. It shows how a region of high 
$\mcM^{\langle \cdot \rangle}_\mcQ$ appears as the coupling increases, as long
as the detuning is not too strong ($\sim 10$). This is because large couplings
induce a rapid decay in the quantum capacity while the oscillations from the 
detuning restore the capacity. For large detunings, the probability for
transitions in the atom is low, which implies that the capacity initially has
small oscillations close to one with and a slow decay.
For small couplings $\gamma/\lambda<1/2$ and zero detuning, the capacity decays
monotonically \cite{sabrinaScienRepot}, this makes all the measures discussed
equal to zero and therefore a useless  \qv.
\begin{table} 
     \begin{ruledtabular}
     \begin{tabular}{cccccc}
     & $\delta/\lambda$ & $\mcM^{\<\cdot\>}_\mcQ$&$\mcM^{\rm max}_\mcQ$ &$\mcM_\mcQ^{\infty}$\\
    \hline
     {\color{black} \LARGE \textbf{---}} & 0 & 0.4072 & 0.6419 & 1.4322     \\
     {\color{blue} \LARGE \textbf{---}} & 40 & 0.4301 & 0.8154 & 4.4928     \\
     {\color{black} \bm{$- - -$}} & 80 & 0.2564 & 0.4963 & 6.0953\\
     {\color{blue} \bm{$- - -$}} & 150 & 0.1210 & 0.2309 & 4.7588
     \end{tabular}
     \end{ruledtabular}
     \caption{The table shows the measures treated in this work of the quantum
     capacities shown in figure \ref{acap}, with $t_\cut{}=\infty$.
     \label{tableap}}
\end{table} 
\section{Conclusions} 
In the light of considerable advances in the experimental manipulation of 
quantum systems at the very fundamental level, understanding and controlling 
how a quantum system interacts with its surroundings is of paramount 
importance. In this context finite, structure rich environments play an 
important role, and the challenge has been to understand and control the 
resulting non-Markovian evolution.   
In particular, one might wonder whether there is a possibility to
take advantage of the flow of information back to the system which is 
characteristic to non-Markovianity. Defining and quantifying non-Markovianity is
a non-trivial task. In this work we have shown that one can
define and quantify non-Markovian behavior in a physically meaningful way. 
One that is insightful and avoids 
the drawbacks of previous attempts, like divergence in very generic cases, and 
counter intuitive outcomes. Moreover we could define the new measure with a 
task in mind: hiding and retrieving classical or quantum information using
a quantum channel. The efficiency with which this taks is accomplished, is 
directly related to the NM measure. Finally, we have illustrated the proposed
measures with simple physical examples.\\

\begin{acknowledgments}
We acknowledge support from the grants UNAM-PAPIIT IN111015,
and CONACyT 153190 and 129309, as well as useful discussions with Sabrina 
Maniscalco and Heinz-Peter Breuer. I.G.M. and D.A.W. received support from 
ANPCyT (Grant No.PICT 2010-1556), UBACyT, and CONICET 
(Grants No. PIP 114-20110100048 and No. PIP 11220080100728). I.G.M and C.P. 
share a bi-national grant from CONICET (Grant No. MX/12/02 Argentina) and 
CONACYT (Mexico). 
\end{acknowledgments}


\begin{thebibliography}{35}%
\makeatletter
\providecommand \@ifxundefined [1]{%
 \@ifx{#1\undefined}
}%
\providecommand \@ifnum [1]{%
 \ifnum #1\expandafter \@firstoftwo
 \else \expandafter \@secondoftwo
 \fi
}%
\providecommand \@ifx [1]{%
 \ifx #1\expandafter \@firstoftwo
 \else \expandafter \@secondoftwo
 \fi
}%
\providecommand \natexlab [1]{#1}%
\providecommand \enquote  [1]{``#1''}%
\providecommand \bibnamefont  [1]{#1}%
\providecommand \bibfnamefont [1]{#1}%
\providecommand \citenamefont [1]{#1}%
\providecommand \href@noop [0]{\@secondoftwo}%
\providecommand \href [0]{\begingroup \@sanitize@url \@href}%
\providecommand \@href[1]{\@@startlink{#1}\@@href}%
\providecommand \@@href[1]{\endgroup#1\@@endlink}%
\providecommand \@sanitize@url [0]{\catcode `\\12\catcode `\$12\catcode
  `\&12\catcode `\#12\catcode `\^12\catcode `\_12\catcode `\%12\relax}%
\providecommand \@@startlink[1]{}%
\providecommand \@@endlink[0]{}%
\providecommand \url  [0]{\begingroup\@sanitize@url \@url }%
\providecommand \@url [1]{\endgroup\@href {#1}{\urlprefix }}%
\providecommand \urlprefix  [0]{URL }%
\providecommand \Eprint [0]{\href }%
\providecommand \doibase [0]{http://dx.doi.org/}%
\providecommand \selectlanguage [0]{\@gobble}%
\providecommand \bibinfo  [0]{\@secondoftwo}%
\providecommand \bibfield  [0]{\@secondoftwo}%
\providecommand \translation [1]{[#1]}%
\providecommand \BibitemOpen [0]{}%
\providecommand \bibitemStop [0]{}%
\providecommand \bibitemNoStop [0]{.\EOS\space}%
\providecommand \EOS [0]{\spacefactor3000\relax}%
\providecommand \BibitemShut  [1]{\csname bibitem#1\endcsname}%
\let\auto@bib@innerbib\@empty
\bibitem [{\citenamefont {Breuer}\ and\ \citenamefont
  {Petruccione}(2007)}]{breuer2007theory}%
  \BibitemOpen
  \bibfield  {author} {\bibinfo {author} {\bibfnamefont {H.}~\bibnamefont
  {Breuer}}\ and\ \bibinfo {author} {\bibfnamefont {F.}~\bibnamefont
  {Petruccione}},\ }\href {http://books.google.com.mx/books?id=DkcJPwAACAAJ}
  {\emph {\bibinfo {title} {The Theory of Open Quantum Systems}}}\ (\bibinfo
  {publisher} {OUP Oxford},\ \bibinfo {year} {2007})\BibitemShut {NoStop}%
\bibitem [{\citenamefont {Schlosshauer}(2007)}]{decoherencetheory}%
  \BibitemOpen
  \bibfield  {author} {\bibinfo {author} {\bibfnamefont {M.~A.}\ \bibnamefont
  {Schlosshauer}},\ }\href {http://books.google.com.ar/books?id=DkcJPwAACAAJ}
  {\emph {\bibinfo {title} {Decoherence and the Quantum-To-Classical
  Transition}}}\ (\bibinfo  {publisher} {Springer},\ \bibinfo {year}
  {2007})\BibitemShut {NoStop}%
\bibitem [{\citenamefont {Raimond}\ \emph {et~al.}(2001)\citenamefont
  {Raimond}, \citenamefont {Brune},\ and\ \citenamefont
  {Haroche}}]{Haroche2001}%
  \BibitemOpen
  \bibfield  {author} {\bibinfo {author} {\bibfnamefont {J.~M.}\ \bibnamefont
  {Raimond}}, \bibinfo {author} {\bibfnamefont {M.}~\bibnamefont {Brune}}, \
  and\ \bibinfo {author} {\bibfnamefont {S.}~\bibnamefont {Haroche}},\ }\href
  {\doibase 10.1103/RevModPhys.73.565} {\bibfield  {journal} {\bibinfo
  {journal} {Rev. Mod. Phys.}\ }\textbf {\bibinfo {volume} {73}},\ \bibinfo
  {pages} {565} (\bibinfo {year} {2001})}\BibitemShut {NoStop}%
\bibitem [{\citenamefont {Leibfried}\ \emph {et~al.}(2003)\citenamefont
  {Leibfried}, \citenamefont {Blatt}, \citenamefont {Monroe},\ and\
  \citenamefont {Wineland}}]{Blatt2003}%
  \BibitemOpen
  \bibfield  {author} {\bibinfo {author} {\bibfnamefont {D.}~\bibnamefont
  {Leibfried}}, \bibinfo {author} {\bibfnamefont {R.}~\bibnamefont {Blatt}},
  \bibinfo {author} {\bibfnamefont {C.}~\bibnamefont {Monroe}}, \ and\ \bibinfo
  {author} {\bibfnamefont {D.}~\bibnamefont {Wineland}},\ }\href {\doibase
  10.1103/RevModPhys.75.281} {\bibfield  {journal} {\bibinfo  {journal} {Rev.
  Mod. Phys.}\ }\textbf {\bibinfo {volume} {75}},\ \bibinfo {pages} {281}
  (\bibinfo {year} {2003})}\BibitemShut {NoStop}%
\bibitem [{\citenamefont {Bloch}\ \emph {et~al.}(2008)\citenamefont {Bloch},
  \citenamefont {Dalibard},\ and\ \citenamefont {Zwerger}}]{Bloch2008}%
  \BibitemOpen
  \bibfield  {author} {\bibinfo {author} {\bibfnamefont {I.}~\bibnamefont
  {Bloch}}, \bibinfo {author} {\bibfnamefont {J.}~\bibnamefont {Dalibard}}, \
  and\ \bibinfo {author} {\bibfnamefont {W.}~\bibnamefont {Zwerger}},\
  }\href@noop {} {\bibfield  {journal} {\bibinfo  {journal} {Rev. Mod. Phys.}\
  }\textbf {\bibinfo {volume} {80}},\ \bibinfo {pages} {885} (\bibinfo {year}
  {2008})}\BibitemShut {NoStop}%
\bibitem [{\citenamefont {Furusawa}\ and\ \citenamefont
  {Vuckovic}(2009)}]{photon-tech}%
  \BibitemOpen
  \bibfield  {author} {\bibinfo {author} {\bibfnamefont {A.}~\bibnamefont
  {Furusawa}}\ and\ \bibinfo {author} {\bibfnamefont {J.}~\bibnamefont
  {Vuckovic}},\ }\href@noop {} {\bibfield  {journal} {\bibinfo  {journal} {Nat.
  Photonics}\ }\textbf {\bibinfo {volume} {3}},\ \bibinfo {pages} {160402}
  (\bibinfo {year} {2009})}\BibitemShut {NoStop}%
\bibitem [{\citenamefont {Ladd}\ \emph {et~al.}(2010)\citenamefont {Ladd},
  \citenamefont {Jelezko}, \citenamefont {Laflamme}, \citenamefont {Nakamura},
  \citenamefont {Monroe},\ and\ \citenamefont {O'Brien}}]{Ladd:2010gb}%
  \BibitemOpen
  \bibfield  {author} {\bibinfo {author} {\bibfnamefont {T.~D.}\ \bibnamefont
  {Ladd}}, \bibinfo {author} {\bibfnamefont {F.}~\bibnamefont {Jelezko}},
  \bibinfo {author} {\bibfnamefont {R.}~\bibnamefont {Laflamme}}, \bibinfo
  {author} {\bibfnamefont {Y.}~\bibnamefont {Nakamura}}, \bibinfo {author}
  {\bibfnamefont {C.}~\bibnamefont {Monroe}}, \ and\ \bibinfo {author}
  {\bibfnamefont {J.~L.}\ \bibnamefont {O'Brien}},\ }\href@noop {} {\bibfield
  {journal} {\bibinfo  {journal} {Nature}\ }\textbf {\bibinfo {volume} {464}},\
  \bibinfo {pages} {45} (\bibinfo {year} {2010})}\BibitemShut {NoStop}%
\bibitem [{\citenamefont {Lindblad}(1976)}]{Lin76}%
  \BibitemOpen
  \bibfield  {author} {\bibinfo {author} {\bibfnamefont {G.}~\bibnamefont
  {Lindblad}},\ }\href@noop {} {\bibfield  {journal} {\bibinfo  {journal}
  {Commun. Math. Phys.}\ }\textbf {\bibinfo {volume} {48}},\ \bibinfo {pages}
  {119} (\bibinfo {year} {1976})}\BibitemShut {NoStop}%
\bibitem [{\citenamefont {Gorini}\ \emph {et~al.}(1976)\citenamefont {Gorini},
  \citenamefont {Kossakowski},\ and\ \citenamefont {Sudarshan}}]{GoKoSu76}%
  \BibitemOpen
  \bibfield  {author} {\bibinfo {author} {\bibfnamefont {V.}~\bibnamefont
  {Gorini}}, \bibinfo {author} {\bibfnamefont {A.}~\bibnamefont {Kossakowski}},
  \ and\ \bibinfo {author} {\bibfnamefont {E.~C.~G.}\ \bibnamefont
  {Sudarshan}},\ }\href {\doibase http://dx.doi.org/10.1063/1.522979}
  {\bibfield  {journal} {\bibinfo  {journal} {J. Math. Phys}\ }\textbf
  {\bibinfo {volume} {17}},\ \bibinfo {pages} {821} (\bibinfo {year}
  {1976})}\BibitemShut {NoStop}%
\bibitem [{\citenamefont {Breuer}(2012)}]{Breuer12}%
  \BibitemOpen
  \bibfield  {author} {\bibinfo {author} {\bibfnamefont {H.-P.}\ \bibnamefont
  {Breuer}},\ }\href {http://stacks.iop.org/0953-4075/45/i=15/a=154001}
  {\bibfield  {journal} {\bibinfo  {journal} {J. Phys. B: At. Mol. Opt. Phys}\
  }\textbf {\bibinfo {volume} {45}},\ \bibinfo {pages} {154001} (\bibinfo
  {year} {2012})}\BibitemShut {NoStop}%
\bibitem [{\citenamefont {Rivas}\ \emph {et~al.}(2014)\citenamefont {Rivas},
  \citenamefont {Huelga},\ and\ \citenamefont {Plenio}}]{RHP14}%
  \BibitemOpen
  \bibfield  {author} {\bibinfo {author} {\bibfnamefont {A.}~\bibnamefont
  {Rivas}}, \bibinfo {author} {\bibfnamefont {S.~F.}\ \bibnamefont {Huelga}}, \
  and\ \bibinfo {author} {\bibfnamefont {M.~B.}\ \bibnamefont {Plenio}},\
  }\href@noop {} {\bibfield  {journal} {\bibinfo  {journal} {Rep. Prog. Phys.}\
  }\textbf {\bibinfo {volume} {77}},\ \bibinfo {pages} {094001} (\bibinfo
  {year} {2014})}\BibitemShut {NoStop}%
\bibitem [{\citenamefont {Huelga}\ \emph {et~al.}(2012)\citenamefont {Huelga},
  \citenamefont {Rivas},\ and\ \citenamefont
  {Plenio}}]{PhysRevLett.108.160402}%
  \BibitemOpen
  \bibfield  {author} {\bibinfo {author} {\bibfnamefont {S.~F.}\ \bibnamefont
  {Huelga}}, \bibinfo {author} {\bibfnamefont {A.}~\bibnamefont {Rivas}}, \
  and\ \bibinfo {author} {\bibfnamefont {M.~B.}\ \bibnamefont {Plenio}},\
  }\href {\doibase 10.1103/PhysRevLett.108.160402} {\bibfield  {journal}
  {\bibinfo  {journal} {Phys. Rev. Lett.}\ }\textbf {\bibinfo {volume} {108}},\
  \bibinfo {pages} {160402} (\bibinfo {year} {2012})}\BibitemShut {NoStop}%
\bibitem [{\citenamefont {Cormick}\ \emph {et~al.}(2013)\citenamefont
  {Cormick}, \citenamefont {Bermudez}, \citenamefont {Huelga},\ and\
  \citenamefont {Plenio}}]{Cormick2013}%
  \BibitemOpen
  \bibfield  {author} {\bibinfo {author} {\bibfnamefont {C.}~\bibnamefont
  {Cormick}}, \bibinfo {author} {\bibfnamefont {A.}~\bibnamefont {Bermudez}},
  \bibinfo {author} {\bibfnamefont {S.~F.}\ \bibnamefont {Huelga}}, \ and\
  \bibinfo {author} {\bibfnamefont {M.~B.}\ \bibnamefont {Plenio}},\ }\href
  {http://stacks.iop.org/1367-2630/15/i=7/a=073027} {\bibfield  {journal}
  {\bibinfo  {journal} {New J. Phys.}\ }\textbf {\bibinfo {volume} {15}},\
  \bibinfo {pages} {073027} (\bibinfo {year} {2013})}\BibitemShut {NoStop}%
\bibitem [{\citenamefont {Reich}\ \emph {et~al.}()\citenamefont {Reich},
  \citenamefont {Katz},\ and\ \citenamefont {Koch}}]{nm-qc}%
  \BibitemOpen
  \bibfield  {author} {\bibinfo {author} {\bibfnamefont {D.~M.}\ \bibnamefont
  {Reich}}, \bibinfo {author} {\bibfnamefont {N.}~\bibnamefont {Katz}}, \ and\
  \bibinfo {author} {\bibfnamefont {C.~P.}\ \bibnamefont {Koch}},\ }\href@noop
  {} {\bibinfo  {journal} {arxiv:1409.7497}\ }\BibitemShut {NoStop}%
\bibitem [{\citenamefont {Deffner}\ and\ \citenamefont
  {Lutz}(2013)}]{DeffnerLutz2013}%
  \BibitemOpen
\bibfield  {journal} {  }\bibfield  {author} {\bibinfo {author} {\bibfnamefont
  {S.}~\bibnamefont {Deffner}}\ and\ \bibinfo {author} {\bibfnamefont
  {E.}~\bibnamefont {Lutz}},\ }\href {\doibase 10.1103/PhysRevLett.111.010402}
  {\bibfield  {journal} {\bibinfo  {journal} {Phys. Rev. Lett.}\ }\textbf
  {\bibinfo {volume} {111}},\ \bibinfo {pages} {010402} (\bibinfo {year}
  {2013})}\BibitemShut {NoStop}%
\bibitem [{\citenamefont {Breuer}\ \emph {et~al.}(2009)\citenamefont {Breuer},
  \citenamefont {Laine},\ and\ \citenamefont {Piilo}}]{PhysRevLett.103.210401}%
  \BibitemOpen
  \bibfield  {author} {\bibinfo {author} {\bibfnamefont {H.-P.}\ \bibnamefont
  {Breuer}}, \bibinfo {author} {\bibfnamefont {E.-M.}\ \bibnamefont {Laine}}, \
  and\ \bibinfo {author} {\bibfnamefont {J.}~\bibnamefont {Piilo}},\ }\href
  {\doibase 10.1103/PhysRevLett.103.210401} {\bibfield  {journal} {\bibinfo
  {journal} {Phys. Rev. Lett.}\ }\textbf {\bibinfo {volume} {103}},\ \bibinfo
  {pages} {210401} (\bibinfo {year} {2009})}\BibitemShut {NoStop}%
\bibitem [{\citenamefont {Rivas}\ \emph {et~al.}(2010)\citenamefont {Rivas},
  \citenamefont {Huelga},\ and\ \citenamefont {Plenio}}]{Rivas2010}%
  \BibitemOpen
  \bibfield  {author} {\bibinfo {author} {\bibfnamefont {A.}~\bibnamefont
  {Rivas}}, \bibinfo {author} {\bibfnamefont {S.}~\bibnamefont {Huelga}}, \
  and\ \bibinfo {author} {\bibfnamefont {M.}~\bibnamefont {Plenio}},\ }\href
  {\doibase 10.1103/PhysRevLett.105.050403} {\bibfield  {journal} {\bibinfo
  {journal} {Phys. Rev. Lett.}\ }\textbf {\bibinfo {volume} {105}},\ \bibinfo
  {pages} {050403} (\bibinfo {year} {2010})}\BibitemShut {NoStop}%
\bibitem [{\citenamefont {Lu}\ \emph {et~al.}(2010)\citenamefont {Lu},
  \citenamefont {Wang},\ and\ \citenamefont {Sun}}]{Xiao2010}%
  \BibitemOpen
  \bibfield  {author} {\bibinfo {author} {\bibfnamefont {X.-M.}\ \bibnamefont
  {Lu}}, \bibinfo {author} {\bibfnamefont {X.}~\bibnamefont {Wang}}, \ and\
  \bibinfo {author} {\bibfnamefont {C.~P.}\ \bibnamefont {Sun}},\ }\href
  {\doibase 10.1103/PhysRevA.82.042103} {\bibfield  {journal} {\bibinfo
  {journal} {Phys. Rev. A}\ }\textbf {\bibinfo {volume} {82}},\ \bibinfo
  {pages} {042103} (\bibinfo {year} {2010})}\BibitemShut {NoStop}%
\bibitem [{\citenamefont {Vasile}\ \emph {et~al.}(2011)\citenamefont {Vasile},
  \citenamefont {Maniscalco}, \citenamefont {Paris}, \citenamefont {Breuer},\
  and\ \citenamefont {Piilo}}]{VasileManisc2011}%
  \BibitemOpen
  \bibfield  {author} {\bibinfo {author} {\bibfnamefont {R.}~\bibnamefont
  {Vasile}}, \bibinfo {author} {\bibfnamefont {S.}~\bibnamefont {Maniscalco}},
  \bibinfo {author} {\bibfnamefont {M.~G.~A.}\ \bibnamefont {Paris}}, \bibinfo
  {author} {\bibfnamefont {H.-P.}\ \bibnamefont {Breuer}}, \ and\ \bibinfo
  {author} {\bibfnamefont {J.}~\bibnamefont {Piilo}},\ }\href {\doibase
  10.1103/PhysRevA.84.052118} {\bibfield  {journal} {\bibinfo  {journal} {Phys.
  Rev. A}\ }\textbf {\bibinfo {volume} {84}},\ \bibinfo {pages} {052118}
  (\bibinfo {year} {2011})}\BibitemShut {NoStop}%
\bibitem [{\citenamefont {Luo}\ \emph {et~al.}(2012)\citenamefont {Luo},
  \citenamefont {Fu},\ and\ \citenamefont {Song}}]{Shunlong2012}%
  \BibitemOpen
  \bibfield  {author} {\bibinfo {author} {\bibfnamefont {S.}~\bibnamefont
  {Luo}}, \bibinfo {author} {\bibfnamefont {S.}~\bibnamefont {Fu}}, \ and\
  \bibinfo {author} {\bibfnamefont {H.}~\bibnamefont {Song}},\ }\href {\doibase
  10.1103/PhysRevA.86.044101} {\bibfield  {journal} {\bibinfo  {journal} {Phys.
  Rev. A}\ }\textbf {\bibinfo {volume} {86}},\ \bibinfo {pages} {044101}
  (\bibinfo {year} {2012})}\BibitemShut {NoStop}%
\bibitem [{\citenamefont {Lorenzo}\ \emph {et~al.}(2013)\citenamefont
  {Lorenzo}, \citenamefont {Plastina},\ and\ \citenamefont
  {Paternostro}}]{Lorenzo2013}%
  \BibitemOpen
  \bibfield  {author} {\bibinfo {author} {\bibfnamefont {S.}~\bibnamefont
  {Lorenzo}}, \bibinfo {author} {\bibfnamefont {F.}~\bibnamefont {Plastina}}, \
  and\ \bibinfo {author} {\bibfnamefont {M.}~\bibnamefont {Paternostro}},\
  }\href {\doibase 10.1103/PhysRevA.88.020102} {\bibfield  {journal} {\bibinfo
  {journal} {Phys. Rev. A}\ }\textbf {\bibinfo {volume} {88}},\ \bibinfo
  {pages} {020102} (\bibinfo {year} {2013})}\BibitemShut {NoStop}%
\bibitem [{\citenamefont {Bylicka}\ \emph {et~al.}(2014)\citenamefont
  {Bylicka}, \citenamefont {Chru\'{s}ci\'{n}ski},\ and\ \citenamefont
  {Maniscalco}}]{sabrinaScienRepot}%
  \BibitemOpen
  \bibfield  {author} {\bibinfo {author} {\bibfnamefont {B.}~\bibnamefont
  {Bylicka}}, \bibinfo {author} {\bibfnamefont {D.}~\bibnamefont
  {Chru\'{s}ci\'{n}ski}}, \ and\ \bibinfo {author} {\bibnamefont
  {Maniscalco}},\ }\href@noop {} {\bibfield  {journal} {\bibinfo  {journal}
  {Sci. Rep.}\ }\textbf {\bibinfo {volume} {4}},\ \bibinfo {pages} {5720}
  (\bibinfo {year} {2014})}\BibitemShut {NoStop}%
\bibitem [{\citenamefont {\ifmmode \check{Z}\else
  \v{Z}\fi{}nidari\ifmmode~\check{c}\else \v{c}\fi{}}\ \emph
  {et~al.}(2011)\citenamefont {\ifmmode \check{Z}\else
  \v{Z}\fi{}nidari\ifmmode~\check{c}\else \v{c}\fi{}}, \citenamefont {Pineda},\
  and\ \citenamefont {Garc\'ia-Mata}}]{Znidaric2011}%
  \BibitemOpen
  \bibfield  {author} {\bibinfo {author} {\bibfnamefont {M.}~\bibnamefont
  {\ifmmode \check{Z}\else \v{Z}\fi{}nidari\ifmmode~\check{c}\else
  \v{c}\fi{}}}, \bibinfo {author} {\bibfnamefont {C.}~\bibnamefont {Pineda}}, \
  and\ \bibinfo {author} {\bibfnamefont {I.}~\bibnamefont {Garc\'ia-Mata}},\
  }\href@noop {} {\bibfield  {journal} {\bibinfo  {journal} {Phys. Rev. Lett.}\
  }\textbf {\bibinfo {volume} {107}},\ \bibinfo {pages} {080404} (\bibinfo
  {year} {2011})}\BibitemShut {NoStop}%
\bibitem [{\citenamefont {Divincenzo}\ \emph {et~al.}(2002)\citenamefont
  {Divincenzo}, \citenamefont {Leung},\ and\ \citenamefont
  {Terhal}}]{Divincenzo2002}%
  \BibitemOpen
  \bibfield  {author} {\bibinfo {author} {\bibfnamefont {D.~P.}\ \bibnamefont
  {Divincenzo}}, \bibinfo {author} {\bibfnamefont {D.~W.}\ \bibnamefont
  {Leung}}, \ and\ \bibinfo {author} {\bibfnamefont {B.~M.}\ \bibnamefont
  {Terhal}},\ }\href@noop {} {\bibfield  {journal} {\bibinfo  {journal} {IEEE
  Trans. Inf. Theory}\ }\textbf {\bibinfo {volume} {48}},\ \bibinfo {pages}
  {580} (\bibinfo {year} {2002})}\BibitemShut {NoStop}%
\bibitem [{\citenamefont {DiVincenzo}\ \emph {et~al.}(2003)\citenamefont
  {DiVincenzo}, \citenamefont {Hayden},\ and\ \citenamefont
  {Terhal}}]{Divincenzo2003}%
  \BibitemOpen
  \bibfield  {author} {\bibinfo {author} {\bibfnamefont {D.}~\bibnamefont
  {DiVincenzo}}, \bibinfo {author} {\bibfnamefont {P.}~\bibnamefont {Hayden}},
  \ and\ \bibinfo {author} {\bibfnamefont {B.}~\bibnamefont {Terhal}},\ }\href
  {\doibase 10.1023/A:1026013201376} {\bibfield  {journal} {\bibinfo  {journal}
  {Found. Phys}\ }\textbf {\bibinfo {volume} {33}},\ \bibinfo {pages} {1629}
  (\bibinfo {year} {2003})}\BibitemShut {NoStop}%
\bibitem [{\citenamefont {Matthews}\ \emph {et~al.}(2009)\citenamefont
  {Matthews}, \citenamefont {Wehner},\ and\ \citenamefont
  {Winter}}]{Mathews2009}%
  \BibitemOpen
  \bibfield  {author} {\bibinfo {author} {\bibfnamefont {W.}~\bibnamefont
  {Matthews}}, \bibinfo {author} {\bibfnamefont {S.}~\bibnamefont {Wehner}}, \
  and\ \bibinfo {author} {\bibfnamefont {A.}~\bibnamefont {Winter}},\ }\href
  {\doibase 10.1007/s00220-009-0890-5} {\bibfield  {journal} {\bibinfo
  {journal} {Commun. Math. Phys.}\ }\textbf {\bibinfo {volume} {291}},\
  \bibinfo {pages} {813} (\bibinfo {year} {2009})}\BibitemShut {NoStop}%
\bibitem [{\citenamefont {Jaynes}\ and\ \citenamefont
  {Cummings}(1963)}]{JaynesCumm1963}%
  \BibitemOpen
  \bibfield  {author} {\bibinfo {author} {\bibfnamefont {E.~T.}\ \bibnamefont
  {Jaynes}}\ and\ \bibinfo {author} {\bibfnamefont {F.~W.}\ \bibnamefont
  {Cummings}},\ }\href@noop {} {\bibfield  {journal} {\bibinfo  {journal}
  {Proc. IEEE}\ }\textbf {\bibinfo {volume} {51}},\ \bibinfo {pages} {89}
  (\bibinfo {year} {1963})}\BibitemShut {NoStop}%
\bibitem [{\citenamefont {Bengtsson}\ and\ \citenamefont
  {Zyczkowski}(2006)}]{BenZyc06}%
  \BibitemOpen
  \bibfield  {author} {\bibinfo {author} {\bibfnamefont {I.}~\bibnamefont
  {Bengtsson}}\ and\ \bibinfo {author} {\bibfnamefont {K.}~\bibnamefont
  {Zyczkowski}},\ }\href@noop {} {\emph {\bibinfo {title} {Geometry of Quantum
  States: An Introduction to Quantum Entanglement}}}\ (\bibinfo  {publisher}
  {Cambridge University Press},\ \bibinfo {year} {2006})\BibitemShut {NoStop}%
\bibitem [{\citenamefont {Heinosaari}\ and\ \citenamefont
  {Ziman}(2012)}]{HeiZim12}%
  \BibitemOpen
  \bibfield  {author} {\bibinfo {author} {\bibfnamefont {T.}~\bibnamefont
  {Heinosaari}}\ and\ \bibinfo {author} {\bibfnamefont {M.}~\bibnamefont
  {Ziman}},\ }\href@noop {} {\emph {\bibinfo {title} {The mathematical language
  of quantum theory: From uncertainty to entanglement}}}\ (\bibinfo
  {publisher} {Cambridge University Press},\ \bibinfo {year}
  {2012})\BibitemShut {NoStop}%
\bibitem [{\citenamefont {Smith}()}]{Smith2010}%
  \BibitemOpen
  \bibfield  {author} {\bibinfo {author} {\bibfnamefont {G.}~\bibnamefont
  {Smith}},\ }\href@noop {} {\bibinfo  {journal} {arxiv:1007.2855}\
  }\BibitemShut {NoStop}%
\bibitem [{\citenamefont {Garc\'ia-Mata}\ \emph {et~al.}(2012)\citenamefont
  {Garc\'ia-Mata}, \citenamefont {Pineda},\ and\ \citenamefont
  {Wisniacki}}]{NachoCarlosDiego2012}%
  \BibitemOpen
\bibfield  {journal} {  }\bibfield  {author} {\bibinfo {author} {\bibfnamefont
  {I.}~\bibnamefont {Garc\'ia-Mata}}, \bibinfo {author} {\bibfnamefont
  {C.}~\bibnamefont {Pineda}}, \ and\ \bibinfo {author} {\bibfnamefont
  {D.}~\bibnamefont {Wisniacki}},\ }\href {\doibase 10.1103/PhysRevA.86.022114}
  {\bibfield  {journal} {\bibinfo  {journal} {Phys. Rev. A}\ }\textbf {\bibinfo
  {volume} {86}},\ \bibinfo {pages} {022114} (\bibinfo {year}
  {2012})}\BibitemShut {NoStop}%
\bibitem [{\citenamefont {Garc\'ia-Mata}\ \emph {et~al.}(2014)\citenamefont
  {Garc\'ia-Mata}, \citenamefont {Pineda},\ and\ \citenamefont
  {Wisniacki}}]{NachoCarlosDiego2014}%
  \BibitemOpen
  \bibfield  {author} {\bibinfo {author} {\bibfnamefont {I.}~\bibnamefont
  {Garc\'ia-Mata}}, \bibinfo {author} {\bibfnamefont {C.}~\bibnamefont
  {Pineda}}, \ and\ \bibinfo {author} {\bibfnamefont {D.~A.}\ \bibnamefont
  {Wisniacki}},\ }\href {http://stacks.iop.org/1751-8121/47/i=11/a=115301}
  {\bibfield  {journal} {\bibinfo  {journal} {J. Phys. A: Math. Theor.}\
  }\textbf {\bibinfo {volume} {47}},\ \bibinfo {pages} {115301} (\bibinfo
  {year} {2014})}\BibitemShut {NoStop}%
\bibitem [{\citenamefont {Leboeuf}\ \emph {et~al.}(1990)\citenamefont
  {Leboeuf}, \citenamefont {Kurchan}, \citenamefont {Feingold},\ and\
  \citenamefont {Arovas}}]{leboeuf1990}%
  \BibitemOpen
  \bibfield  {author} {\bibinfo {author} {\bibfnamefont {P.}~\bibnamefont
  {Leboeuf}}, \bibinfo {author} {\bibfnamefont {J.}~\bibnamefont {Kurchan}},
  \bibinfo {author} {\bibfnamefont {M.}~\bibnamefont {Feingold}}, \ and\
  \bibinfo {author} {\bibfnamefont {D.}~\bibnamefont {Arovas}},\ }\href@noop {}
  {\bibfield  {journal} {\bibinfo  {journal} {Phys.Rev. Lett.}\ }\textbf
  {\bibinfo {volume} {65}},\ \bibinfo {pages} {3076} (\bibinfo {year}
  {1990})}\BibitemShut {NoStop}%
\bibitem [{Note1()}]{Note1}%
  \BibitemOpen
  \bibinfo {note} {The case without detuning is treated in detail in~\cite
  {breuer2007theory}; the present case with detuning has first been solved in
  Ref.~\cite {sabrinaScienRepot} but with a minor mistake. Here, we present the
  corrected expression.}\BibitemShut {Stop}%
\bibitem [{\citenamefont {Giovannetti}\ and\ \citenamefont
  {Fazio}(2005)}]{Giovannetti2005}%
  \BibitemOpen
  \bibfield  {author} {\bibinfo {author} {\bibfnamefont {V.}~\bibnamefont
  {Giovannetti}}\ and\ \bibinfo {author} {\bibfnamefont {R.}~\bibnamefont
  {Fazio}},\ }\href {\doibase 10.1103/PhysRevA.71.032314} {\bibfield  {journal}
  {\bibinfo  {journal} {Phys. Rev. A}\ }\textbf {\bibinfo {volume} {71}},\
  \bibinfo {pages} {032314} (\bibinfo {year} {2005})}\BibitemShut {NoStop}%
\end{thebibliography}
%
\end{document}